# QUERY ROUTING AND PROCESSING IN PEER-TO-PEER DATA SHARING SYSTEMS

Raddad Al King, Abdelkader Hameurlain, Franck Morvan

Institut de Recherche en Informatique de Toulouse (IRIT), Université Paul Sabatier
118, route de Narbonne, F-31062 Toulouse Cedex 9, France
E-mail: {alking, hameur, morvan}@irit.fr

## ABSTRACT

*Sharing musical files via the Internet was the essential motivation of early P2P systems. Despite of the great success of the P2P file sharing systems, these systems support only "simple" queries. The focus in such systems is how to carry out an efficient query routing in order to find the nodes storing a desired file. Recently, several research works have been made to extend P2P systems to be able to share data having a fine granularity (i.e. atomic attribute) and to process queries written with a highly expressive language (i.e. SQL). These works have led to the emergence of P2P data sharing systems that represent a new generation of P2P systems and, on the other hand, a next stage in a long period of the database research area.*

*The characteristics of P2P systems (e.g. large-scale, node autonomy and instability) make impractical to have a global catalog that represents often an essential component in traditional database systems. Usually, such a catalog stores information about data, schemas and data sources. Query routing and processing are two problems affected by the absence of a global catalog. Locating relevant data sources and generating a close to optimal execution plan become more difficult. In this paper, we concentrate our study on proposed solutions for the both problems. Furthermore, selected case studies of main P2P data sharing systems are analyzed and compared.*

## 1. INTRODUCTION

Nowadays, Peer-to-Peer (hereafter P2P) systems become very popular. This popularity can be seen as a result of the features of these systems such as: scalability, node autonomy, self-configuration and decentralized control. P2P systems offer a good opportunity to overcome the limitations of the Client/Server based systems. By avoiding bottlenecks and being fault tolerant, P2P systems are suitable for large-scale distributed environments in which nodes (interchangeably called peers) can share their resources (e.g. computing power, storage capacity, network bandwidth) in an autonomously and decentralized manner. The more the resources are available in a P2P system, the more the computing power and the storage capacity have important values. This advantage enables P2P systems to perform complex tasks with relatively low cost without any need to powerful servers. In the next section, we highlight the notion of "*P2P Systems*".

### 1.1. P2P Systems

There is no agreement about what are P2P systems. Through our reading, we find several definitions of these systems [40, 52, 57]. The definition of [52] represents the systems having one or more servers while the definition of [49] ignores this type of systems. Thus, we agree with the definition of Milojicic et al. [52] as "*The term "peer-to-peer" (P2P) refers to a class of systems and applications that employ distributed resources to perform a function in a decentralized manner. The resources encompass computing power, data (storage and content), network bandwidth, and presence (computers, human, and other resources). The critical function can be distributed computing, data/content sharing, communication and collaboration, or platform services. Decentralized may apply to algorithms, data, and metadata or to all of them*". Even if there is no standard definition of P2P systems, most researchers characterize them by: (i) scalability in terms of the node number and the resource number; (ii) node autonomy; (iii) dynamicity, (iv) resource heterogeneity, (v) decentralized control and (vi) self-configuration. In such systems, each node can act as: (i) a server when it offers its resources to be used by other nodes, (ii) client when it uses the resources of other nodes, (iii) a router when it propagates coming queries and messages to other nodes and (iv) data source[1] when it shares its own data with the system nodes. The researches on P2P systems become more and more numerous and the contexts in which we use these systems become also too much numerous. In this paper, we focus our study on the P2P database context.

## 1.2. P2P Systems and Database Systems

P2P systems are successfully used in several domains such as: file sharing, computing power sharing and instant message exchange. Due to their "good" features, new domains aim to take advantage of these systems. In the public health domain, for instance, we can cite some examples: (i) a doctor in a hospital may want to share most of his own data with other colleagues and to hide a portion of his data for personal reasons (e.g. data of an experience concerning a new drug for Alzheimer's disease); (ii) a doctor treating ill person may want to access the databases of the family doctor and the pharmacy of his patient in order to know his medical history and (iii) several researchers around the world are working on a drug for Alzheimer's disease want to share data stored in their databases during an experience.

Traditional database systems are not able to answer the requirements of the previous examples. Distributed database systems (DDBS) are utilized when data is fragmented on many sites, known a priori, and the administration is centred on one control site. DDBS can mange a few dozens of databases [61]. Data warehousing systems require transporting all data available on data sources to one control site. This approach could create a bottleneck especially when the data source number is big and the dynamicity of the system is high. Data integration systems are largely used nowadays to integrate virtually data stored in data sources distributed on the Internet. However, a data integration system could manage just a few hundreds of data sources [61]. We believe that P2P data sharing systems can play an important role in the examples cited above. A P2P data sharing system could be seen as "*a large-scale distributed system, in which, the nodes are autonomous and can join and leave the system in a completely decentralized manner. Each node has its own database system composing of a DBMS and one or more database(s) that it manages*". However, there is no scientific consensus on how to design and how to implement this type of systems. That is because of many issues such as architecture, data representation, security, query processing and optimization still considered as open problems. We next highlight the problematic addressed in this paper.

## 1.3. Query Routing and Processing

File sharing is the most popular application of nowadays P2P systems. In a file sharing system, given a query, the system finds nodes storing the desired file and returns their IP addresses to the user in order to select a node for downloading the file. Usually, the semantics of the file name is known by the majority of the users. This knowledge is obtained via the media (e.g. TV) or through the user social environment (e.g. university, school). The focus of a file sharing query processing is how to route efficiently the query to relevant data sources. In order to share data with fine granularity, P2P systems must provide more database functionality such as schema matching and query optimization. Despite of many research efforts that have been done in this direction, processing queries written with a highly expressive language (i.e. SQL) is still challenging.

Due to the dynamicity and the large scale of P2P data sharing systems, it is not practical to have a global catalog witch represents an important component of traditional database systems. The authors of [34] cite some types of information (called metadata) stored in a global catalog: (i) Information about database schemas (e.g. table definitions, views, integrity constraints); (ii) Information concerning partitioning schema such as "*what global tables have been partitioned and how they can be reconstructed*" [34] and (iii) Physical information such as the placement of partitioned data and the statistics that are utilized to calculate the cost function of an execution plan. The statistics could be: (i) parameters concerning physical characteristics of data (e.g. relation size), (ii) physical parameters about data sources (e.g. CPU load), and (iii) physical parameters concerning the network (e.g. network bandwidth).

---

[1] A data source could be a file stored site or a database site.

Locating relevant data sources and carrying out efficient query processing are two open problems affected by the absence of a global catalog. Many solutions have been proposed in order to adapt query types more advanced than file sharing queries. The authors of [16] have offered solutions for processing Range Queries. In this type of queries, the user is interested by a data whose values belonging to a précised range. The authors of [27] propose solutions for processing Aggregation Queries. Solutions for processing Top-k Queries are proposed by the authors of [2]. In this type of queries, the user could be satisfied by k best answers, for a given query, found in the system. As our knowledge, we can mention that the proposed solutions focus on a particular type of queries. There is no solution for processing all or a combination of these types of queries.

### 1.4. Paper Goals and Organization

Considerable amount of surveys has been done in the P2P data sharing field. For instance, Milojicic et al.[40] present an overview of what is "*P2P computing*" in general terms. The authors of [8] distinguish between DB-centric and P2P-centric features of some P2P and distributed database systems. In [57], P2P content distribution models are discussed and compared while XML data management techniques for P2P environments are studied in [33]. Search and security issues are discussed in [14]. In this paper, our objective is to give an overview of selected projects of P2P data sharing systems and to present a qualitative comparison between these projects, on one hand, and our solutions that have been already published [29, 30, 31, 32] on the other hand. Due to the space limitations, we concentrate our study on the three problems: (i) query routing, (ii) schema matching and (iii) query optimization. We believe that these problems are more difficult in P2P data sharing systems than traditional database systems that usually have a global catalog and accept central administration.

The rest of this paper is organized as follows. In section 2, we study the query routing problem while the schema matching one is addressed in section 3. We focus on query optimization issues in section 4. Before the conclusion in section 6, we represent in section 5 some projects of P2P data sharing systems and we compare them with our proposed solutions.

## 2. QUERY ROUTING

In P2P environments, query routing becomes a difficult problem with the absence of a global catalog that holds often information about data placement. The issue here is how to route efficiently a given query to relevant data sources. Nodes in P2P environment form a virtual network (called overlay network) located above a physical network (i.e. the Internet). The topology of a P2P system indicates how its peers are situated on the virtual network. It has a strong impact on the query routing efficiency in terms of the number of exchanged messages, on one hand, and the answer quality on the other hand. Answer quality means the ability to return all valid answers existed in the system. According to their topologies, P2P systems are classified into three main classes:

- **Unstructured P2P systems:** in this class, there are no privileged nodes and all nodes play the same role. The node autonomy is high in the sense that nodes are not forced to be structured according to a predefined geometric shape. Each node establishes a direct connection with one or more nodes called neighbors. Generally, it is not require that a node be aware about the resources available on its neighbors.

- **Structured P2P Systems:** in this class of P2P systems, all nodes must be structured into a specific geometric shape (e.g. ring). They must also know some information about their neighbors and about their contents without be able to choose these neighbors.

- **Super-peer P2P systems:** plying the same role by all nodes in a system could be considered as a strong hypothesis. Practically, nodes in a system have neither the same computing power nor the same storage capacity. So it is better to assign different roles to nodes according to their computing and storage capabilities. For this reason, a hybrid class based on the both paradigms P2P and Client/Server has been proposed. In this class, powerful nodes (called super-peers) play the role of a server while the other nodes act as clients. The super-peers can be organized into structured or unstructured topology. Having several super-peers facilitates shared resource administration. However, the fault tolerance is decreased in the super-peer P2P systems. When a super-peer fails, its clients become isolated from the rest of the system.

We notice that the query routing methods depend often on the P2P classes (unstructured, structured and super-peer). In this section, we discuss an important part of the methods utilized in each one of these classes.

## 2.2. Query Routing in Unstructured P2P Systems

In order to send a query towards relevant data sources, the Peer Initiating the Query (PIQ) sends identical messages to its neighbors. Each message contains the query, the PIQ identifier and a value of the parameter TTL (Time To Live). This parameter represents the maximum number of peers that a message is allowed to pass through. After having received the message, each peer executes the query and sends the answer to the PIQ. Furthermore, it decreases the value of TTL by 1. If the value of TTL becomes 0, this peer destroys the message. Otherwise, the peer sends, in its turn, the message containing the new value of TTL to its neighbors. By continuing this process, all peers situated at a distance having a value lower than the initial value of TTL are going to receive the message. Other peers in the system may not be able to receive the message even if they have valid answers. According to this approach, the peers have no knowledge about the content of their neighbors. For this reason, this approach is called Blind Routing (BR) approach. Even if the BR approach maintains a high degree of peer autonomy, it is very costly in terms of network bandwidth consumption. The BR approach needs a big number of messages to be exchanged during one query routing process. Thus may lead to network flooding especially when the peers' number is very high.

Several solutions have been proposed to improve the performance of the BR approach by storing information about neighbor contents. An iterative method has been proposed in [65]. According to this method instead of having one value of TTL, we choose many values between 1 and $V_{max}$. For each value, we repeat the BR approach. If this method finds an acceptable number of valid answers for a value of TTL lower than $V_{max}$, we stop repeating the BR approach. This method improves the performance of the BR approach in terms of the number of exchanged messages but the response time of this method is higher. In [28], the authors propose to send the message received by a peer to only a part of its neighbors instead of sending the message to all neighbors. The choice of this part is randomly carried out. Then, this solution decreases the number of exchanged messages. However, the fact of not sending the message to all the neighbors allows ignoring a lot of valid answers. More intelligent solution has been proposed in the same paper [28], the basic idea is to use statistics concerning former executed queries. It is true that, according to this solution, the choice of the neighbors is not random. But, it is true also that this new solution requires a big number of messages to update the statistics stored in the system when one peer join/leave the system, on one hand, and when a new query is executed on the other hand.

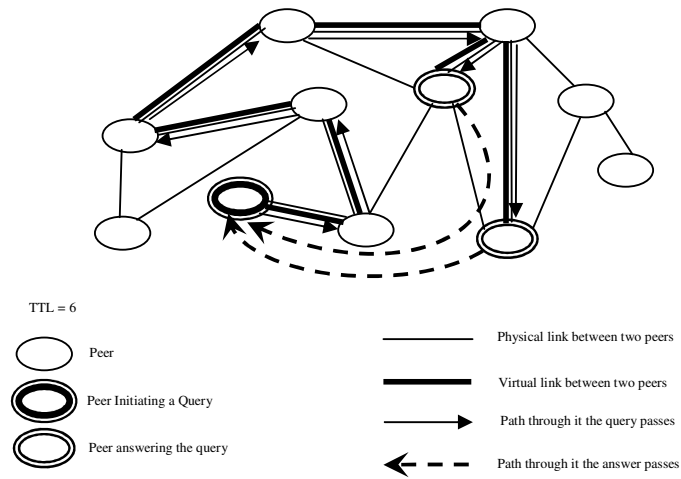

Figure 1. Query Routing in Unstructured P2P Systems

The authors of [51] have proposed a probability based mechanism for choosing the neighbors that must receive the message. Each peer stores probability information about its neighbors' contents. In addition to the ignorance of valid answers, the probability information must be up-to-dated regularly. This is could be expensive when the peers' number is very high and when the peers join and leave regularly the system. The authors of [63] and [12] have proposed the mechanism of Local Indices according to which each peer indexes the data stored in the peers situated on a surface having r as diameter value. This mechanism improves relatively the response time of the BR approach. However, when a peer joins/leaves the system, a flood of messages in a zone (having r peers as diameter) could be emerged. Another solution has been proposed for locating data sources. The principle of this solution is that every peer has to index all data stored in a peer that have already answered a former query. This solution is good and practical for repetitive queries. However, one of its drawbacks is the number of the update messages required when a peer has to leave the system.

Despite of the big efforts that have been done to improve the query routing performance in unstructured P2P systems, the proposed solutions are still incapable to return all valid answers existed in the system.

### 2.2. Query Routing in Structured P2P Systems

In the structured P2P systems, all peers are organized as a geometric shape. The methods utilized to route a query depends on the geometric shape considered by the system. These methods differ from each others in the way in which a peer chooses its neighbors and shares information with them. Thereafter, we discuss some examples of the geometric shapes (Hypercube, Ring, Cartesian space of *d* dimensions and Tree) according to which the peers can be organized. We discuss as well the impacts of these shapes on the query routing process.

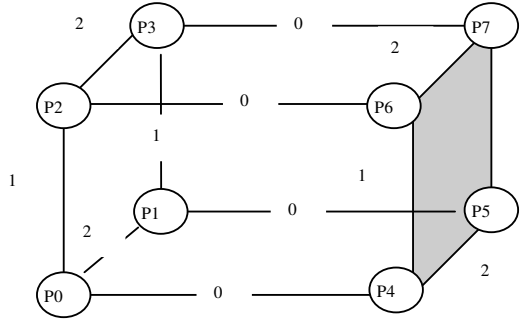

**Figure 2. Example of a Hypercube**

### 2.2.1. Hypercube

As a good example of this topology, we study HyperCup [56]. According to this geometric shape, all peers are organized in a hypercube topology having as diameter $\Delta = \log_b N$ [48], where $b$ is the base of the hypercube (that means that in each dimension there are $b$ peers) and $N$ is the number of all peers in the system. The diameter of a hypercube is defined as "*the shortest path between most distant nodes in terms of node hops*" [56]. The number of dimensions is $(L_{max}+1)$ where $L_{max}$ is done as $N = b^{L_{max}+1}$. Each peer has $(b-1)*(L_{max}+1)$ neighbors. The links between peers are labeled by numbers between 0 and $b$. Thus allows the neighbors to be ordered in a symmetric way. When the link between two peers X and Y is labeled by $i$, it means that X is $i^{th}$ neighbor of Y and verse versa. In figure 2, we present a hypercube having b = 2 as a base. We notice that the diameter is $\Delta = \log_2 8 = 3$ and that each peer has $(2-1)*(2+1) = 3$ neighbors. The query routing in the hypercube topology is made as follows. Firstly, the peer initiating the query broadcast messages to all its neighbors. Each message contains the query and the label value of the link by witch the query passes. When a peer receives a message coming from peer $j$, for instance, it sends the message towards the neighbors whose links are labeled by a number bigger than $j$ and modifies the label value tagged with the message. For example, when peer P0 sends a message to the neighbors P4, P2 and P1, peer P4 sends the query to peers P5 and P6. Peer P1 does not send the query because the links with its neighbors P3 and P5 are labeled by a number lower than 2 which is the label of its link with P0. As for peer P2, it sends the query to peer P3. The third steps, peer P6 sends the message to peer P7. This method of routing allows locating a desired data item by performing $O(\log_b N)$ hops.

### 2.2.2. Ring

According to the ring geometry, all peers must be ordered on a virtual ring. Each peer has an identifier on the ring. Several systems use this geometry as [11, 49, 54, 66]. As an example of this topology, we study Chord protocol [54] which is largely utilized in structured P2P systems. Chord utilizes the DHT (Distributed Hash Table) technology which requires storing shared information in a table fully distributed on all peers. One essential role of the DHT is to guide the query routing in order to avoid network flooding and to guarantee fining all valid answers existed in the system. The information of a DHT is distributed on peers by using a hash function known by all peers. Each data item is represented by a key that is created by using the hash function of the DHT. Each peer stores the pairs (key, identifier) concerning the keys for which it is responsible in its part of the DHT. The keys are distributed in a balanced manner on the peers. The identifier of a peer is represented by $m$ bits and each key is represented also by $m$ bits. The peer responsible for a key $k$, for instance, is the peer having the first identifier that equals or succeeds *k*. Each peer *P* maintains a routing table having *m* entries and the $i^{th}$ entry contains the identifier of the first peer that succeeds *P* by at least $2^{i-1}$ on the virtual ring.

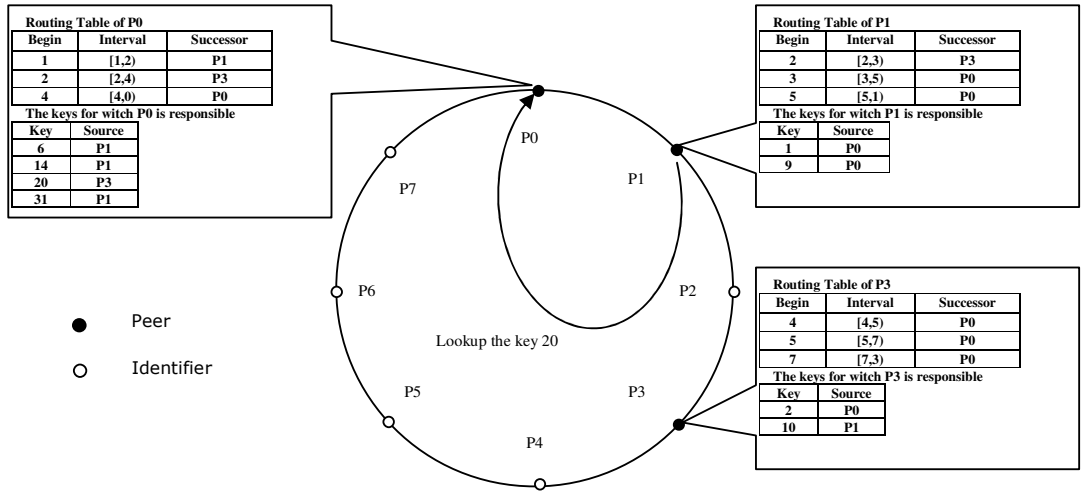

**Figure 3. Example of a Ring**

Chord allows locating the peer responsible for a given key via O(Log N) hops where N is the peers' number. To illustrate the query routing algorithm utilized by Chord, we study a case of a ring with identifiers represented by m = 3 bits. In figure 3, we suppose that peer P1 requires to lookup the key 20. It has to look in its routing table. Instead of looking for the key 20, it looks for the key 4 (because 20 mod (8) = 4). After this consultation, it sends the query to peer P0. The peer P0 knows that the data item represented by the key 20 is stored on peer P3. It can tell P1 of this information. If P1 has asked to obtain the data item represented by the key 20, in this case, P0 sends the message to peer P3 which sends the data item to P1.

### 2.2.3. *d* Dimension Cartesian Space

This topology is based on a Cartesian space of *d* dimensions. This space is independent from the underlying physical network and decomposed into separate zones. Each zone is allocated to one peer. The P2P systems using the CAN (Content Addressable Network) virtual network [48] are based on this topology. In CAN, each peer is responsible for a storage space (its zone). It means that each peer stores all the pairs (key, values) that their keys fall in its zone according to a hash function. Each key is represented by *d* coordinated k= ($c_1$, $c_2$, … $c_d$).

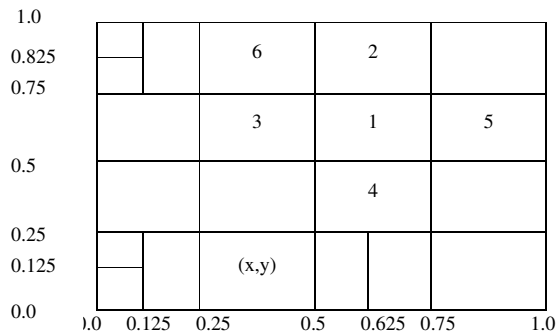

**Figure 4. Example of *d* Dimension Cartesian**

A greedy algorithm is used to route a given query towards the neighbors that their zones are the most close to the zone in which the lookup key fall. For example, in figure 4, we consider a Cartesian space of d=2 dimensions and we suppose that a peer possesses a zone *i* is called *Pi*. Let us suppose that *P1* needs to lookup the key (x, y). This peer sends a message to *P4* because this peer possesses the closest zone, in terms of Euclidian distance, to the zone in witch the key (x, y) falls. Then, peer *P4* sends the message to the neighbor possessing the closest zone to the zone holding (x, y). This process is continued until the message arrives at its destination. According to this topology, each peer has to store information concerning 2*d neighbors and the average number of hops required to lookup a key is O( $dN^{1/d}$ ) where N is the number of all peers in the system.

### 2.2.4. Tree

As an example of this topology, we study the case of Baton [26] which is utilized to create a structure of distributed index by redistributing data on all peers in a balanced manner. Each node in the tree represents a peer in the system. It is identified by the number of its level in the tree and its order in this level. Thus, each peer has a logical identifier corresponding to its level and to its order in this level and a physical identifier corresponding to its IP address. Furthermore, each peer has a link (by using the physical identifiers) with its parent, its left neighbor, its right neighbor, its sons and other selected neighbors in its level. The selected neighbors are maintained in two routing tables, one for the right neighbors and the other for the left neighbors. The number of entries of each table is O(log N) or N is the number of all peers in the system. The $i^{th}$ entry of the right table (the left table) of a peer p contains a link with the peer having the number $n+2^{i-1}$ (respectively $n-2^{i-1}$) at the same level. The data are redistributed on the peers in a way that each peer becomes responsible for a range of values. In figure 5, for example, peer 6 is the parent of peers 12 and 13 then its range ([54, 61)) has to be in the middle of their ranges ([51, 54) and [61, 68) respectively).

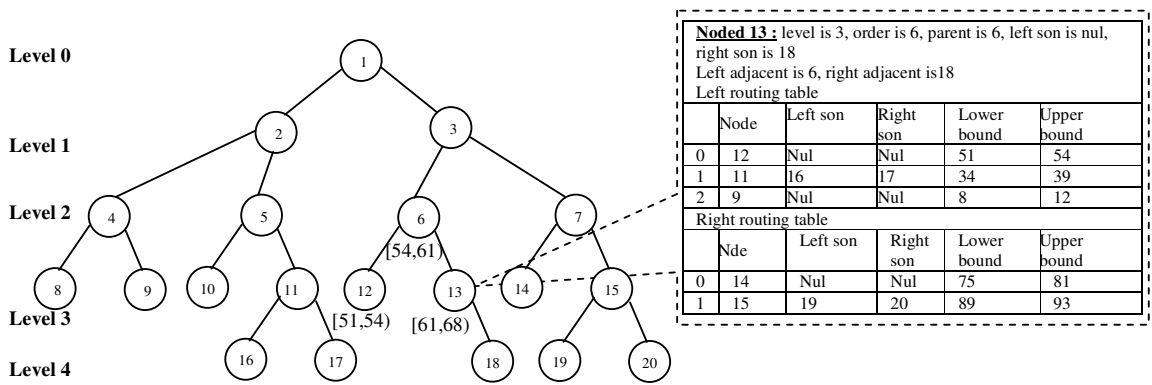

**Figure 5. Example of a Tree**

The routing of an equal query is performed as follow. If the neighbors having the upper bounds (lower bounds) of their smaller (bigger) ranges than those of the query range, the query is sent horizontally towards the neighbor farthest who has an upper bound (lower bound) smaller (bigger) than the values of the query range. Then, the query is vertically propagated by following the left (right) neighbor or left (right) son. This process is repeated until the query arrives on the peer storing the desired data item. The number of hops to be carried out during this query routing is O (LogN). As regards the range queries, the previous process is repeated until we find the peers storing the lower bound (or upper bound) of the query range. Then, it is necessary to look for the rest of the range. For that purpose, the query has to visit x peers. Consequently, the complexity of the query routing becomes O(Log N + x).

### 2.2.5. Discussion

| Topology | Algorithm | Parameters | Query routing complexity | Join/leave complexity | DHT |
|---|---|---|---|---|---|
| Ring | Chord | N is the peers' number | $1/2 Log N$ | $Log^2 N$ | Yes |
| d Dimension Cartesian space | CAN | N is the peers' number d is the number of dimension of CAN | $dN^{1/d}$ | Join: $d/2N^{1/d}$ Leave: $2d$ | Yes |
| Tree | Baton | N is the peers' number x is the number of peers to be contacted after having found the first part of the range asked by the query | $LogN$ and $LogN+x$ for Range Queries | $Log N$ | No |
| Hypercube | HyperCup | N is the peers' number b is the base of the Hypercube | $Log_b N$ | $Log_b N$ | No |

Table 1. Comparison between the different structured P2P systems based on the studied shapes

In the table 1, we compare the four types of structured systems that are studied above. We use the following parameters for the comparison: (i) *query routing complexity* witch represents the average number of peers by which a query has to pass via during its routing. (ii) *join/leave complexity* that represents the number of peers to be contacted when a peer join/leave the system and (iii) *with or without DHT*.

Among the four geometric shapes, the best is Hypercube and the worst is CAN in terms of query routing complexity. Baton does not separate the virtual network of the underlying physical network. Having information about the physical network is not practical because the network parameters are regularly changed. Another drawback of Baton is that only the peers forming leaves of the tree can leave the system at any time. However, if another peer wants to leave the system it must find a leaf in order to exchange their places. We can also add that in the case of Baton, it is necessary to redistribute the data on all the peers according to the values of their data items. Thus is not practical when the volume of shared data is very height. It is more practical to redistribute information concerning the placements of the data items and to maintain storing the data items on their sources as in the case of Chord. Consequently, Baton is suitable for building a fully distributed data index in some applications where we can move the data items. In terms of join/leave the system, Both Baton and Hypercube are more efficient than CAN and Chord. That is due to the fact that Baton and Hypercube have no DHT. The DHT facilitates the query routing and guarantees to find all valid answers existed in the system. A DHT allows a well-balanced distribution of charge and information. It does not require moving data from their sources. CAN and Chord are based on the DHT technology.

### 2.3. Query Routing in Super-Peer P2P Systems

We remind that in the super-peer P2P systems, there are two types of peers: (i) the super-peers who are often powerful peers and (ii) the normal peers, or simply the peers, which are clustered in a way that each cluster is connected to a super-peer. Each super-peer is responsible for indexing the data items stored in the peers of its cluster. The super-peers form a P2P system. This system can be structured or unstructured. We have already noticed that this class of systems is hybrid between both paradigms Client/Server and P2P. In the Super-peer P2P systems, the query routing is done as follows. The peer initiating the query sends a message, containing the query, to its super-peer that looks for relevant data sources in its local index. If the super-peer finds relevant data sources, it answers the PIQ by sending it their IP addresses. If not, the super-peer contacts its neighbors (at the super-peer level) found in the system. Each super-peer looks for relevant data sources in its local index, if it finds valid answers; it sends the message to its clients representing relevant data sources. The clients send their answers to their super-peer which send these answers, in its turn, to the super-peer responsible for the PIQ. Finally, the answers will be sent to the PIQ.

In P2P data sharing systems, efficient query routing is not enough to locate relevant data sources. Schema heterogeneity problem has a direct impact on the data source localization. Because of this problem, a relevant data source couldn't be found during the query routing phase even if it is the only relevant source available in the system. In order to solve this problem, P2P data sharing systems must provide efficient functionalities for matching heterogeneous schemas. Therefore, we next study schema matching issues.

## 3. SCHEMA MATCHING

In P2P data sharing systems, databases are designed and developed autonomously. Each node describes its data using its own local schema. By "schema", we mean a plan (or a shape) that describes the relationships between data items and represents a set of rules defining these relationships. Because of the node autonomy at database design level, the local schemes could be semantically and structurally heterogeneous.

P1:   Doctor (Name, Salary, Phone_number)
P8:   Doc (Name, Paycheck, Telephone)
        Ill (Name, Address, Doctor_Name, SSN)
        Treat (Pressure, Temperature, Drug, Trade_Name, Dose)
P14:  Physician (Name, Address)
        Patient (Name, Age)
P21:  Consultant (Name, Earnings, Address, Telephone)
P32:  Consultant (Name, Salary, Telephone)
P42:  Doctor (Name, Address, Tel)
P48:  Patient (Name, Age, Address, Doctor_Name, Disease,
        Social_Security_Number)
        Treatment (Pressure, Temp, Drug, Trade_name, Dose)

**Figure 6. Local schemas of a set of peers**

In figure 6, we illustrate an example of heterogeneous peer schemas used in the medical domain. We consider relational schemas because of the large use of relational database model. The schema heterogeneity could appear at two levels:

- **Semantic level:** witch involves:
  - Synonymy: it occurs when multiple terms having the same meaning. For instance, "Consultant" in P32 schema and "Doctor" in P1 schema are two terms holding the meaning of "a person training to treat ill people".

  - Polysemy: it occurs when one term having many meanings. For example, "Consultant" on P21 has the meaning of expert which is defined as "a person having special knowledge of a subject, gained as a result of training or experience". It is not the same meaning of "Consultant" on P32.

- **Structural level:** that refers to the heterogeneity of the attributes representing the same relation in several schemas. For instance, Doctor (Name, Salary, Phone_number) according to P1 schema and Doctor (Name, Address, Tel) according to P42 schema are structurally heterogeneous.

When a node submits a query written in terms of its local schema, other nodes must understand the query in order to be able to answer it correctly. The schema matching is the process of finding for each element of a given schema, its correspondent element in another schema and to produce matching rules for reformulating a query written in terms of one schema into the terms of the other schema. In order to do that, this process must exploit information related to both schemas. For instance, the names of schema' elements, their data types and domain ranges.

### 3.1. Schema Matching Types

The schema matching has a significant impact on query processing. If the matching rules are "badly" created, the quality of query answers could be "bad" also. We mean that some answers may be invalid or may not be those that are desired by the peer initiating the query. Thus, important questions may arise at this point such as: (i) how to create the schema matching? (ii) when? and (iii) by who, by experts or by end users? Since it is not possible to carry out the schema matching in a totally dynamic way (without any human intervention) [19], this process can be done manually (without any machine intervention) or semi-automatically. The drawback of manual schema matching is that it is time consuming. Therefore, semi-automatic approaches have emerged as a solution to this drawback. Researchers are still trying to automate, as much as possible, the schema matching process. A summary of the proposed approaches in this area is presented in [51]. We can distinguish between two types of schema matching:

- **Static Matching:** in this matching type, the matching rules are created at the system design time and remain unchanged (except, of course during maintenance). The matching rules can be stored in a centralized manner on one or several sites known by all peers or distributed in any way on all peers. They will be used during a query processing in order to translate a given query between different schemas. These rules are often created by experts in schema designing who are familiar with the semantics of the matched schemas.

- **Dynamic Matching:** when the matching rules are created on the fly, or changed regularly, we call this type of matching by dynamic matching. In this type of matching, we can change the matching rules and/or create new rules anytime. Usually, the matching rules are created at the query run-time.

### 3.2. Schema Matching Approaches

In P2P environments, schema matching approaches can be classified as follows: global schema based approach, pairwise matching approach and information retrieval based approach. We next explain each approach.

- **Global Schema based Approach :** The absence of a global schema in P2P systems can be considered by some researchers as a strong hypothesis. Even if the peer instability and the large scale prevent having information concerning data placement, peers can agree on a global schema. Two approaches LAV (Local-As-View) [37] and GAV (Global - As - View) [41] are existing to create the matching rules between the global schema and the local schemas. According to the LAV approach, the local schemas are defined as views (queries) formulated in terms of the global schema. A query written according to the global schema must be reformulated in terms of the local schemas. On the other hand, in the GAV approach, the global schema is defined as a view on the local schemas. According to this approach, we translate a given query (written in terms of global schema) into sub-requests written according to the local schemas. A hybrid approach GLAV [36] between LAV and GAV approaches can be used.

- **Pairwise Matching Approach:** The use of a global schema could be unacceptable in certain applications for which, it is very difficult to find a common agreement between peers on such a schema. For that reason, certain researchers prefer to use a direct pairwise matching between local schemas. In this way, we create a semantic network whose nodes are the local schemas and the links are the matching rules.

- **Information Retrieval based Approach:** According to this approach, the user has to play an essential role in the matching process by providing keywords. This approach is similar to that utilized in the information search on the Internet by using any search engine. The principle of this approach consists of (i) providing keywords by the user, (ii) finding the sites storing data having semantics close to that of the keywords, (iii) delivering the ID addresses of these sites to the use in order to select the relevant sites and finally (iv) downloading the desired data from the selected sites.

Schema matching defines semantic and structural relationships between heterogeneous schemas. Thus allowing P2P data sharing systems to localize relevant data sources and to process queries similar to that found in P2P file sharing systems by replacing the file name by the name of data item. In order to process more advanced queries, P2P data sharing systems must support other functionalities provided by traditional database systems such as query optimization.

## 4. QUERY OPTIMIZATION

In traditional database systems, when a distributed query processing is carried out, the query is translated by an optimizer into execution plans. An execution plan should define certain details about the following steps: (i) the choice of relevant data sources for each data item, (ii) the order of the query operations and the choice of the more suitable algorithm for each one and (iii) the choice of the target nodes which are able to execute each operation. Then, the optimizer chooses the close to optimal execution plan which minimizes a cost function (e.g. response time). The most an execution plan is close to optimal, the more the query execution is efficient. By "efficient", we mean a minimized value of the cost function. The calculation of the cost function is based on information (usually called metadata) stored in a global catalog. Usually such a catalog contains global information about data, their schemas and their sources.

In traditional database systems, the global catalog storing metadata can be centralized or duplicated on many servers. In P2P data sharing systems, the query optimization is more complex with regard to traditional database systems. The absence of a global catalog in P2P data sharing systems makes creating a close to optimal execution plan a real challenging. Due to the decentralization, the peer autonomy and instability and the large scale of P2P systems, it is not practical to have a centralized catalog which could create a bottleneck. Furthermore, the autonomy of peers and the fact that each peer is a client and a server at the same time, prevent P2P systems to have a catalog duplicated on all peers. This type of catalog requires a big number of update messages when one peer join/leave the system. We can distinguish between two optimization approaches. We next highlight the differences between both approaches.

## 4.1. Centralized Optimization

In this type of query optimization, an execution plan is generated on the peer initiating the query and then it is decomposed into sub-plans that will be sent towards target peers. When a target peer receives a sub-plan, it can re-optimize this sub-plan. But it has no idea about the global execution plan. When a peer submits a query, it has generally neither global vision on its environment nor global information about relevant data sources. It may have information about the resources available on its neighbors only and some statistics coming from previous query executions. The lack of information and the obsolete statistics are two factors on which the optimality of an execution plan depends. Available information and statistics could be obsolete because of the large scale, the node autonomy and instability of P2P environments. Each peer can make updates locally on its own data without informing other peers found in the system or sometimes without being itself connected to the system. Furthermore, the values of some physical parameter concerning target peers (e.g. available memory, CPU load) or those of the network (bandwidth and latency) could be continuously varied because of (i) the enormous number of queries submitted at the same time and (ii) the irregular arriving rate of these queries. So, the generated execution plan can be sub-optimal which could lead to invalid answers.

In our previous paper [29], we have proposed to take advantage of the location phase to obtain all metadata needed for creating a close to optimal execution plan. During the localization phase, relevant data sources are discovered and at the same time metadata concerning the given query are obtained and returned to the PIQ. Even if the optimization process is centralized on the PIQ, this optimization is not satisfied by local metadata available on the PIQ. But it also utilizes metadata available on the data sources. The utilized metadata are fresh and reliable because it is obtained directly from their data sources.

## 4.2. Distributed Optimization

According to this approach, the peer initiating the query generates an initial execution plan by basing on its local metadata. This execution plan will be sent to remote peers that re-optimize it by using their own metadata. Papadimos and Maier have proposed the strategy of Mutant Query Plans (MQP) [46] for sharing XML data. Each peer receives the execution plan with intermediate results coming from another peer. Then, it optimizes again the query by using its local knowledge. Finally, the results with the execution plan will be transmitted to another peer and so on. The MQP strategy has two drawbacks: (i) shipping the intermediate results with the query plan many times for one query witch will be expensive in terms of network bandwidth consumption and (ii) the user query may be optimized and executed on a peer without a prior knowledge about the capability of this peer at the moment of receiving the user query to optimize and/or to execute the query.

A similar strategy to MQP has been proposed by the authors of [38] who distinguish their strategy from MQP by the consideration of the query decomposition and the using of dynamic execution plan based on the cost of the transmission of intermediate results and on the up-to-date information of the neighbors. However, these authors mention, in the same paper [38], that in order to avoid transmitted a big amount of intermediate results, their strategy requires about O(N) messages that must be exchanged in the network where N is the number of all peers in the system.

Even if query optimization is still an open problem in P2P environments, little of research works take attention about this problem. For instance, when we analyze selected projects of P2P data sharing systems in the next section, this problem is not addressed in the majority of the case studies.

## 5. CASE STUDIES IN P2P DATA SHARING SYSTEMS

Recently, several projects for designing and developing P2P data sharing systems have emerged. These projects differ in their methods of: query routing, schema matching, and query optimization. In this section, we discuss several case studies representing essential steps toward P2P data sharing systems. We note that the performance and the services of P2P systems depend often on the system topology. Except the APPA system [2] which represents an exception by being topology independent P2P system.

**APPA** [2] is a P2P system based on a fully distributed platform that is adaptable with different P2P typologies (structured, unstructured or super-peer). The peers in APPA have views on a Common Description Schema (hereafter CSD). This is similar to LAV approach except that the queries in APPA are written in terms of local views and not in terms of the CSD. Each peer stores locally matching rules between its schema and the CSD. Currently, the APPA authors propose solutions for top-k query processing and for the reconciliation of replicated data.

We next give an overview on other case studies of P2P data sharing systems classified as the P2P topology utilized by these systems.

### 5.1. Unstructured P2P Data Sharing Systems

We begin our study by unstructured P2P data sharing systems in witch peers have equivalent role and they don't be forced to be organized on a predefined geometric shape.

**HePTex** [7] is a P2P system for sharing data stored in heterogeneous XML databases. The focus of HePTex is how inferring automatically precise mapping rules from informal schema correspondences and how to translate XQuery queries between the peers' schemas. The authors of HePTex mention that their translation algorithm is the first one that "deals with schema mappings, including data-metadata along and against the direction of mappings". Query routing in HePTex is totally decentralized and has logarithmic complexity (as in DHT-based systems). However, HePTex supposes that each peer has some knowledge about other peers' schemas and the user of this peer must "easily" create correspondences between schemas.

**GrouPeer** [35] is unstructured P2P system that allows sharing relational data by using semantic overlay clusters. The authors of GrouPeer focus on the problem of the inability of a peer to obtain information about desired data or about peers having similar interests. When pairwise matching rules are already established, a peer receiving rewritten query cannot be able to sufficiently answer the query. A given version of rewritten query could be corrupted during the rewriting process on intermediate peers. GrouPeer offers a dynamic approach for creating and maintaining semantic groups of peers. Through this approach, a peer can choose a version of rewritten query or automatically writes its version. The authors investigate the notion of semantic query similarity that decreases when the number of attributes increases.

**OntoZilla** [25] is a P2P system avoiding network flooding by using ontologies to cluster peers into interest groups in order to route queries to relevant peers. Ontologies allow the description of the peer resources which allows automatic information processing. OntoZilla is more flexible than DHT-based systems since the peers' clusters are based on peers' special interests that could be varied over the time. The peers in a cluster utilize only one classification system for special interests, services and information. Currently, the authors develop a prototype and, as our knowledge, there is no experimental information about the validity of OntoZilla system available yet.

### 5.2. Structured P2P Data Sharing Systems

In this sub-section, we study main structured P2P data sharing systems by beginning with our proposed one.

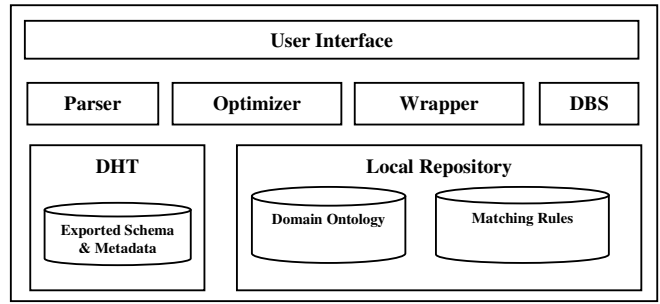

**Figure 7. Layered software architecture on each node of the OP2DBS**

**OP2DBS** [31] is an ontology-based P2P database system for processing SQL based queries. In this system, a query written in terms of a local schema is processed as follows. After parsing the query, (i) Query Reformulation phase is carried out in order to make query understood by remote peers. So, the semantic and structural conflicts must be resolved in this phase. The system provides semi-automatic tools to user in order to reformulate locally the query into another query written on terms of a domain ontology (hereafter DO). After that, (ii) Query Localization and Metadata Obtaining phase localizes relevant data sources. This phase is based on a DHT, it obtains all metadata required for the next phase. (iii) Global optimization phase is similar to global optimization in data integration systems. The only different is that in P2P environments, with the absence of a global catalog, global optimization is based on information obtained during the localization phase. The output is sub-queries written in terms of the DO and sent to remote peers in order to enter in a (iv) Local Optimization phase and then in the (v) Execution phase. Finally the results must be assembled and delivered to the user.

**eSciGrid** [53] is a P2P-based system for sharing huge amount of data in a Grid environment. It allows e-science communities to establish decentralized and cooperative data sharing infrastructures. This system provides a decentralized protocol for caching range query results. This protocol takes into account the network traffic and the physical distance between peers. eSciGrid must be extended to be used in other domains such as e-Health and Astrophysics.

**PIER** [20] is an Internet-scale relational query engine. It supposes the existence of a standard schema replicated on all peers. But, it doesn't suppose the existence of a global catalog. It obtains meta-data on the fly when needed by using a kind of monitoring service. PIER utilizes CAN protocol [48] in which a relation key is constructed from the relation namespace and the resource identifier (resource ID). The relation resource ID is its primary key. The join between two relations requires the multicast of the join query to all peers in the two namespaces attributed to these relations. Instead of using persistent storage, PIER considers soft-state storage in which a temporary data item still alive for a given life-time and then it is discarded.

**PGrid** [3] is a dynamic binary search based P2P system. The originality of this system is due to the separation between peer identifier and peer position on the network (i.e. its IP address). PGrid uses pairwise matching approach and assumes the existence of predefined RDF matching rules between different schemas in the system. These rules are initially created by skilled experts. PGrid is able to extract new matching rules between two local schemas that are not initially matched. It uses routing strategy based on gossiping algorithm that is not affected the system scalability. PGrid proposes a solution for processing RDF-based range queries.

**Pins** [62] is a DHT-based P2P systems for data indexation and interrogation. It stores any kind of objects (e.g. images, videos, documents, etc.). It supports storing also meta-data (i.e. Nom_attribute=valeur) to facilitate data source search. The originality of Pins lives in the fact that it can be used with any DHT-based P2P system and with any type of data. This system supports several services such as the evaluation of declarative queries [62] and the personalized indexation [62]. However, Pins suggest replicating shared objects (data and their meta-data on several peers. When the number of these objects is very high, the management of their replicas becomes very difficult. The authors of Pins [62] do not address the heterogeneousness between the objects stored in Pines. Their future works will be dedicated to add of semantics to the queries processed by the system and to supports various levels of query expressions.

## 5.3. Super-peer P2P Data Sharing Systems

We next study selected P2P data sharing systems that are largely utilized nowadays.

**SIGMCC** [13] is a super-peer P2P system for sharing meta patient records that can be stored by using different identifiers and various structures in many data sources. SIGMCC has a filtering module allowing sending personal data only to authorized clients. It provides XML-based meta-Electronic Patient Record (meta-EPR) that contains information extracted by different EPRs. However, using a specific EPR is relevant to limited applications that have specific requirements.

**Piazza** [22, 58] is super-peer system [8] allows sharing XML data. Each XML schema defines the structure and the content of one peer. Piazza has centralized index storing summaries of the peers' data at different granularity. In addition, the index stores pairwise matching rules representing the relationships between different heterogeneous XML schemas. Thus makes Piazza very similar to a search engine. Piazza users can write their queries according to their local schemas. The Xquery reformulation is the main problem addressed by the Piazza authors. These authors don't explain how can process the queries efficiently. They propose a query reformulation algorithm to obtain all the answers in the existing peers. However, this algorithm is centralized. Moreover, to process a query, a chain of reformulations may be needed. The Piazza authors mention that their reformulation algorithm "*may produce more reformulations than necessary*" [22].

**coDB** [15] is a P2P database system in witch databases are interconnected by GLAV [36] coordination rules. Each peer can be queried, in its local schema, for data that can be fetched from its neighbors if a coordination rule is existed. The architecture of coDB is based on JXTA platform2 which responsible for the peer's activities on the network. Super-peers in coDB can change dynamically the network topology at runtime. Furthermore, a super-peer offers services to start-up all peers, to establish coordination rules, to run experiments and to collect statistical information. Each peer stores a statistical report and makes it available for the user. A user on a peer can show the interactions with the neighbors (called acquaintances) and also the discovered peers (not neighbors).

**Edutella** [43] is a super-peer P2P based on the JXTA platform. It uses RDF for resource notation (including data). Each resource is identified by unique identifier. All notations are represented by a triplet <subject, property, value>, where subject identifies the resource we want to describe (using URI), property indicates which attribute is specified and value indicates the value of the specified attribute. The descriptions of a group of peers are stored on its super-peer. An RDF Schema (RDFS) is used. This schema contains resource classes, properties and conditions on the properties (domain, range, etc.). The properties can be used to integrate two schemas, to link several resources or to create hierarchical relationships between resources. Edutella supports RDF-based Top-k query processing.

**PeerDB** system [42, 44] allows sharing relational data by using Information Retrieval approach and mobile agents to find relevant data sources. Mobile agents participate also in the query processing. The overly network is composed of simple peers and LIGLO (Location Independent Global Names Lookup) servers. LIGLO are used to enhance the query routing by assigning unique IDs to the peers and maintaining traces of their current status (connected or disconnected). For each schema item, the user creates keywords (i.e. synonyms). For a given SQL query, mobile agents are created and propagated in the system. The mobile agents use Matching Strategy in order to return information (IP addresses, semantic mappings) about peers that are probably relevant data sources. Then the user chooses the "best" data sources before creating other mobile agents for processing the submitted query. One of the drawbacks of PeerDB is that there is neither global schema nor common semantics about shared data. The connections between user keywords can lead to false query reformulation and then to invalid query answers. Moreover, the PeerDB authors don't address the problem of generation a global execution plan. They use statistics stored in LIGLO servers for the aim of query routing optimization and not for global query optimization purpose.

**AmbientDB** [10] is a prototype of a P2P database management system. It consists of the following components: (i) Relational query processor that is based on a global known schema, (ii) P2P networking protocol which manages the interactions between the peers by basing on a DHT (iii) XML schema integrator which is responsible for solving the heterogeneity between local schemas by using XML files and (iv) Local database which is managed by a relational DBMS. To process a query, the authors assume the existence of an execution plan but they say nothing about the information on which this plan is based, and without saying anything about how to generate this plan. They propose some opportunities for optimizing their algorithms in terms of network bandwidth consumption. But there are no concrete results available at the moment of writing this paper.

### 5.4. Qualitative Comparison

In this section, we present a qualitative comparison between, on one hand, P2P data sharing systems [2, 3, 7, 10, 13, 15, 20, 25, 35, 42, 43, 53, 58, 62] studied in the previous section and the system OP2DBS considered in our research works presented in the papers [29, 30, 31, 32].

---

[2] https://jxta.dev.java.net/

HePTex [7], GrouPeer [35] and OntoZilla [25] are unstructured P2P systems that maintain a high node autonomy level and a good fault tolerance. However, despite the big efforts that have been made they are often incapable to find all valid answers found in the system during the query routing process.

coDB [15], Edutella [43], Piazza [58], PeerDB [42] and SIGMCC [13] are super-peer P2P systems. In such systems, peers do not play equivalent roles. The Piazza System [58] utilizes a central index installed on one (or multiple) node(s) for storing information needed in the query reformulation phase. The system PeerDB utilizes LIGLO servers [42] to improve the query routing quality. As for the coDB system [15], it stores matching rules between heterogeneous schemas on the super-peers. The system Edutella [43] utilizes several super-peers to enhance the data source localization, on one hand, and to execute queries on the other hand. Despite the reduced number of exchanged messages during the data source localization, these systems are less fault tolerant compared to other P2P systems (structured and unstructured). Thus, the ability of these systems to find all valid answers depends on the type of P2P system formed by the super-peers.

AmbientDB [10], PGrid [3], PIER [20], PinS [62], eSciGrid [53] and our system are structured P2P systems capable to provide solutions to the limitations of unstructured and super-peer P2P systems. One advantage of structured P2P systems is the ability to find all valid answers available in the system while using an "acceptable" number of inter-nodes exchanged messages. The structured P2P systems suffer from the fact that the node autonomy is limited comparing with unstructured and super-peers P2P systems. It is true that the nodes can not choose their neighbors, but this can be tolerated when we look at the benefits of these systems. PIER [20], Pins [62] and our system are DHT based P2P systems. The PGrid [3] system distinguishes itself from other structured P2P systems by the separation between the node identifier and its position on the physical network (i.e. its IP address). Moreover, in the PGrid system [3], several nodes may be responsible for storing meta-data concerning a shared data item. Thus improves the fault tolerance of the PGrid system. However, managing the join/leave of a peer is more complicated. The authors of PGrid [3] do not provide information on the complexity of these two processes. A DHT can facilitate the query routing process while maintaining an "acceptable" complexity to the peer join/leave. The PIER system [20] is based on the CAN protocol [48] which is less efficient in terms of the number of inter-nodes exchanged messages than the both protocols Pastry [49] and Chord [54] that are utilized by PinS [62] and our system respectively. The PinS system [62] is characterized by the fact that it can be used with any routing protocol and it is able to utilize any hash function. Although, even if our system is currently using the protocol Chord [54], it is also independent of the routing protocol and of any hash function. eSciGrid [53] utilizes a protocol taking into account the network traffic and the physical distance between peers. Despite of this fact, the scalability of the system is maintained and the query routing still having logarithmic complexity in terms of the peers' number.

As for the schema heterogeneity problem, the systems Piazza [58], coDB [15], GrouPeer [35], OntoZilla [25], PGrid [3] and HePTox [7] utilize the approach of pairwise matching that may requires a chain of matching process when locating the data sources of a given query. The systems APPA [2] and PIER [20] utilize the global schema based approach for carrying out the schema matching process. This solution is very similar to that used in the data integration systems except in the APPA system [2] queries are written by local schemas. In order to use a global schema, it must predict all nodes can connect to the system and all local schemas must be known before sharing any data in the system. Having a global schema makes difficult for a new node to connect to the system because the user of that node may not

International Journal of Database Management Systems ( IJDMS ), Vol.2, No.2, May 2010

necessarily have the capacity to understand the semantics of the terms used in the global schema. In our system, to create the matching rules between the local schemas, we use an approach hybrid between the approach global schema based matching and that based on information retrieval (IR) which is utilized by both systems PeerDB [42] and Edutella [43]. What distinguishes our system is that we replace the global schema by a domain ontology in which the semantics is explicitly explained. This makes easier for a new peer to connect to the system because the user can understand the semantics of the terms utilized by the ontology and can then create the matching between the ontology and its local schema. Another advantage of using a domain ontology is that thousands of domain ontologies are actually available on the Internet which makes our choice practical. The schema matching in the Edutella [43] system is static process. However, in the PeerDB system [42] and in our system the schema matching is a dynamic process and well adapted with the dynamic nature of P2P environments. At that point, the difference between our system and the system PeerDB [42] is that PeerDB utilizes mobile agents to locate data sources while performing the schema matching at the same time. Mobile agents can visit hundreds of peers and perform the matching between the schemas of these nodes and the local schema of the node initiating the query without prior knowledge of the contents of the visited nodes. Therefore, mobile agents can visit several nodes by consuming more resources (e.g. network bandwidth) without any benefit. However, we use in our system a domain ontology and we provide local matching between the local schema of the node initiating the query and the domain ontology. Even if the system PeerDB utilizes LIGLO [42] servers to guide (based on some statistics) the routing of its agents, our system based on a DHT still more efficient. The authors of PinS [62] are currently silent about the heterogeneity between the objects stored in PinS. They will add semantics to the queries processed by PinS in their future research works. The authors of eSciGrid [53] don't currently address the heterogeneity problem to the best of our knowledge.

The global query optimization becomes a major problem when there is no global catalog in the system. The global catalog contains often metadata needed for generating an execution plan close to optimal. To our knowledge, none of the systems studied in the previous section address, in details, this problem. The systems Edutella [43], coDB [15], AmbientDB [10] and APPA [2] (in his super-peer architecture) utilize information available on the super-peers. The Piazza system [58] utilizes information stored in its central index that can create a bottleneck. Regarding our system, we know that it is not possible to obtain up-to-dated global information (i.e. concerning all nodes in the system). We believe that it is not necessary to have global information to process a given query. Information about data sources is sufficient to generate an execution plan close to optimal. The generated plan will be executed on the data sources. Therefore, we have proposed in our previous paper [29] to obtain all the required meta-data during the data source localization phase.

|  | Topology | Schema Matching |  |
|---|---|---|---|

| | | **Approach** | **Type** | |
|---|---|---|---|---|
| **APPA** | Adapted with any P2P topology | Global Schema based | Static | NA* |
| **PIER** | Structured | Global Schema based | Static | Decentralized Based on monitoring meta-data |
| **AmbientDB** | Structured | Global Schema based | Static | NA |
| **Piazza** | Super-peer [BCO+08] | Pairwise based | Static | NA |
| **coDB** | Super-peer | Pairwise based | Dynamic | NA |
| **PGrid** | Structured | Pairwise based | Static | NA |
| **PeerDB** | Super-peer | User-keyword based | Dynamic | NA |
| **Edutella** | Super-peer | User-keyword based | Static | Centralized Based on meta-data available on super-peers |
| **PinS** | Structured | NA | NA | NA |
| **GrouPeer** | Unstructured | Pairwise based | Dynamic | NA |
| **OntoZilla** | Unstructured | Pairwise based | Dynamic | NA |
| **HePTex** | Unstructured | Pairwise based | Dynamic | NA |
| **SIGMCC** | Super-peer | Global Schema based (Schema is replaced by meta-EPR) | Static | NA |
| **eSciGrid** | Structured | NA | NA | NA |
| **OP2DBS** | Structured | User-keyword based + Global Schema based (Schema is replaced by | Dynamic | Centralized Based on meta-data available on data sources |

* **NA**: Not Addressed at the best of our knowledge

**Table 2. Qualitative comparison between the studied P2P data sharing systems**

## 6. CONCLUSION

The success of the P2P file sharing systems has led to the emergence of P2P data sharing systems that support sharing data having granularity finer than a file. Providing highly robust functionality for large-scale dynamic P2P environments is still a challenge. In this paper, we have addressed query routing, schema matching and query optimization problems in P2P data sharing systems. Furthermore, main research works towards the implementation of this type of systems are analyzed and compared.

International Journal of Database Management Systems ( IJDMS ) , Vol.2, No.2, May 2010[4]     K. Aberer, P. C. Mauroux and M. Hauswirth. "A Framework for Semantic Gossiping". ACM SIGMOD (SIGMOD'02) Record 31(4), pp. 48-53, 2002.

[5]     A. Andrzejak and Z. Xu. "Scalable, efficient range queries for grid information services". In Proc. of the IEEE Int. Conf. on P2P computing, pp : 33-40, 2002.

[6]     R. Blanco, N. Ahmed, D. Hadaller, L.G.A. Sung, H. Li and M.A. Soliman. "A survey of data management in peer-to-peer systems". Technical Report CS-2006- 18, University of Waterloo, 2006.

[7]     A. Bonifati, E. Chang, T. Ho, L. V. Lakshmanan, R. Pottinger and Y. Chung, "Schema mapping and query translation in heterogeneous P2P XML databases". The VLDB Journal, V(19), N(2), pp. 231-256, April 2010.

[8]     A. Bonifati, P. K. Chrysanthis, A. M. Ouksel and K. U. Sattler. "Distributed databases and Peer-to-Peer Databases: Past and Present". SIGMOD Rec. 37, 1, pp. 5-11, March 2008.

[9]     I. Brunkhorst1, H. Dhraief, A. Kemper, W. Nejdl and C. Wiesner. "Distributed Queries and Query Optimization in Schema-Based P2P-Systems". In Proc. of the Int. Workshop on Databases, Information Systems, and Peer-to-Peer Computing (DISP2P'03), pp. 184-199, 2003.

[10]    P. Boncz and C. Treijtel. "AmbientDB: Relational Query Processing in a P2P Network". Int. Workshop on Databases, Information Systems and Peer-to-Peer Computing (DBISP2P), 2003.

[11]    A. Crainiceanu, P. Linga, A. Machanavajjhala, J. Gehrke and J. Shanmugasundaram. "P-ring: an efficient and robust P2P range index structure". In Proc. of the 2007 ACM SIGMOD int. Conf. on Management of Data (SIGMOD '07), pp : 223-234, 2007.

[12]    A. Crespo, and H. G. Molina. "Routing indices for peer-to-peer systems". IEEE Int. Conf. on Distributed Computing Systems" (ICDCS'02), pp : 23-33, 2002.

[13]    M. Cannataro, D. Talia, G. Tradigo, P. Trunfio and P. Veltri, "SIGMCC: A system for sharing meta patient records in a Peer-to-Peer environment". Journal of Future Generation Computer Systems, V(24), N(3), pp.222-234, March 2008.

[14]    N. Daswani, H. G. Molina and B. Yang. "Open problems in data-sharing peer-to-peer systems". Int. Conf. on Database Theory (ICDT'03), pp. 1-15, 2003.

[15]    E. Franconi, G. Kuper, A. Lopatenko, I. Zaiharayeu, "Queries and Updates in the coDB Peer to Peer Database System", in the proceedings of the 30th VLDB, pp. 1277-1280, 2004.

[16]    A. Gupta, D. Agrawal and A. El Abbadi. "Approximate range selection queries in peer-to-peer systems". First Biennial Conference on Innovative Data Systems Research (CIDR'03), pp. 141-151, 2003.

[17]    S. Girdzijauskas, A., Datta and K. Aberer, "Structured overlay for heterogeneous environments: Design and evaluation of oscar". ACM Trans. Auton. Adapt. Syst. 5, 1, pp. 1-25, 2010.

[18]    F. Giunchiglia and I. Zaihrayeu. "Implementing Database Coordination in P2P Networks". The Second Workshop on Semantics in Peer-to-Peer and Grid Computing (SemPGrid'04), 2004.

[19]    Halevy, A. "Why Your Data Won't Mix: Semantic Heterogeneity". ACM Queue 3(8), pp. 50–58, 2005.

[20]    R. Huebsch, B. Chun, J. M. Hellerstein, B. T. Loo, P. Maniatis, T. Roscoe, S. Shenker, I. Stoica and A. R. Yumerefendi. "The Architecture of PIER: an Internet-Scale Query Processor". In Proc. of the 2005 CIDR Conf. pp : 28-43, 2005.

[21]    M. Harren, J.M. Hellerstein, R. Huebsch, B.T. Loo, S. Shenker and I. Stoica. "Complex queries in DHT-based peer-to-peer networks". Int. Workshop on Peerto-Peer Systems (IPTPS'02), pp. 242-259, 2002.

[22]    A. Halevy, Z. Ives, P. Mork and I. Tatarinov. "Piazza: data management infrastructure for semantic web applications". Int. Conf. on World Wide Web (WWW'03), pp : 556-567, 2003.

[23]    A. Ismail, M., Quafafou, G., Nachouki, and M. Hajjar, "Efficient super-peer-based queries routing". MEDES '09. ACM, pp. 91-98, 2009.

[24]    A. Ismail, M. Quafafou, G. Nachouki and M. Hajjar, "Data mining effect in peer-to-peer queries routing". In Proceedings of the International Conference on Management of Emergent Digital Ecosystems (MEDES '09), ACM, pp.65-72, 2009.
136

**Authors**

Raddad AL KING received his PhD degree in computer sciences from Paul Sabatier University, Toulouse, France, in 2010. He is currently a Temporary Research Assistant at the Institute of Research in Computer Science of Toulouse (IRIT). His research interests include data source localization, query processing and optimization in Peer-to-Peer environments. 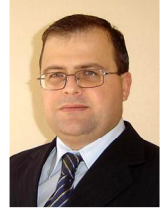

Abdelkader Hameurlain is full professor in Computer Science at Paul Sabatier University, Toulouse, France. He is a member of the Institute of Research in Computer Science of Toulouse (IRIT). His current research interests are in query optimization in parallel and large scale distributed environments, mobile databases, and database performance. Prof. Hameurlain has been the general chair of the International Conference on Database and Expert Systems Applications (DEXA'02). He is co-editor in Chief of the International Journal "Transactions on Large-Scale Data- and Knowledge-Centered Systems" (LNCS, Springer). He was guest editor of two special issues of "International Journal of Computer Systems Science and Engineering on "Mobile Databases" and "Data Management in Grid and P2P Systems". 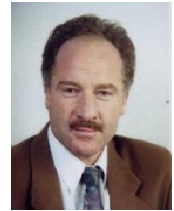

Franck Morvan received a PhD degree in Computer Science from Paul Sabatier University in 1994. He worked at Dassault Data Services society for 3 years before he joined Paul Sabatier University. He is currently associate professor and member of the Institute of Research in Computer Science of Toulouse (IRIT). His main research interests are optimization in distributed and parallel databases, mobile agents and mobile computing.